\documentstyle[prb,twocolumn,aps,floats]{revtex}

\begin{document}

\twocolumn[\hsize\textwidth\columnwidth\hsize\csname
@twocolumnfalse\endcsname

\title{Transport properties of doped $t$-$J$ ladders}

\author{ Hirokazu Tsunetsugu $^{a}$ and Masatoshi Imada $^{b}$}

\address{
$^{a}$Institute of Applied Physics, University of Tsukuba, 
Tsukuba, Ibaraki 305, Japan \\
$^{b}$Institute for Solid State Physics, University of Tokyo, 
Roppongi 7-22-1, Tokyo 106, Japan }

\date{Received}
\maketitle

\begin{abstract}
Conductivity and Hall coefficient for various types of $t$-$J$ ladders 
are calculated as a function of temperature and frequency 
by numerical diagonalization.  
A crossover from an incoherent to a coherent charge dynamics is 
found at a temperature $T_{\rm coh}$.  There exists another crossover
at $T_{\rm PG}$ 
below which a pseudogap opens in the optical spectra, induced by 
the opening of a spin gap.  
In the absence of the spin gap, $T_{\rm coh}$ and the coherent 
weight are suppressed especially with increasing dimensionality.
On the contrary, $T_{\rm coh}$ is strongly enhanced by the pseudogap 
formation below $T_{\rm PG}$, where the coherent Drude weight decreases with 
increasing dimensionality. 
The Hall coefficient shows a strong crossover at $T_{\rm PG}$ below 
which it has large amplitude for small doping concentration.
\end{abstract}

\pacs{PACS numbers:}
\vskip1pc]

In high-$T_c$ cuprates, 
transport phenomena in the normal state show various unusual 
properties in terms of standard normal metals.  
It is well 
known that the in-plane DC conductivity, $\sigma$, is proportional 
to $1/T$ with $T$ being the temperature \cite{rho_T}, and 
the AC conductivity has a broad incoherent part well fitted by 
$\sigma (\omega ) \propto 1 / \omega$ \cite{sigma_omega}. 
Another peculiarity exists 
in the Hall coefficient, $R_H$.  At low $T$'s, 
in spite of the electron density less than half filling, 
$R_H$ is usually positive, meaning hole-like carriers, 
with a large amplitude scaled as $1/\delta$ 
with hole doping $\delta$ \cite{R_delta1,R_delta2}.  
It also shows a crossover to a small amplitude at a rather 
high temperature, $T_{\rm cr}$, and $T_{\rm cr}$ rapidly decreases with 
increasing $\delta$ \cite{Sato,Batlogg}.  
These behaviors are incompatible with 
the standard Fermi liquid theory, implying that 
the carriers change their character with $T$ 
due to strong correlations.  
The incoherent charge dynamics and strongly $T$-dependent 
Hall coefficient have been analyzed in terms of criticality 
near the Mott transition 
together with the effects of preformed pairs \cite{imada}.  
Jakli\v{c} and Prelov\v{s}ek \cite{jaklic} 
calculated $\sigma ( \omega, T )$ 
for the two-dimensional (2D) $t$-$J$ model using 
Lanczos diagonalization.  Their results reproduced the above 
features of the high-$T_c$ cuprates, 
in contrast to the 1D $t$-$J$ model, which is known 
to have a very coherent $\sigma (\omega )$ \cite{horsch}.  

Assaad and Imada showed a large crossover in the $T$-dependence of the 
high-frequency Hall coefficient, $R_H^*$, for the 2D Hubbard model 
using quantum Monte Carlo simulations \cite{assaad}.  
At small $\delta$'s, $R_H^*$ changes its sign twice with decreasing
$T$ from $T$=$\infty$: 
from negative to positive at $T_U$, and then back to negative at 
$T_{\rm AF}$.  In the region of $T_{\rm AF}$$<$$T$$<$$T_U$, 
it is found that strong correlations substantially suppress a quantum 
coherence between different spin configurations in the electron
motion, leading to a large positive Hall coefficient expected for a 
spinless Fermion case.  
Within numerical results, the 2D Hubbard model 
shows no definite indications for preformed pairing or 
superconductivity. It neither reproduces the experimentally 
observed $T$-dependence of $R_H$.  

Recently several two-leg ladder compounds have been experimentally studied.  
A finite spin gap was confirmed in the insulating phases, 
and the superconductivity was reported for doped 
Sr$_{14-x}$Ca$_x$Cu$_{24}$O$_{41}$ under high pressure \cite{uehara}.  
The optical conductivity in the
metallic region shows a similar incoherent character to the high-$T_c$'s
while the resistivity follows $\rho \propto T^2$ at low 
$T$'s in the normal phase \cite{osafune}. 
 
In this paper, we study the conductivity and Hall coefficient of 
the $t$-$J$ two-leg ladders.  
The Hamiltonian of the $t$-$J$ ladder under a magnetic field, $B$, 
perpendicular to the ladder plane is written as 
\begin{eqnarray}
  \label{ham}
  {\cal H} = 
   \sum_{\langle i,j \rangle}
   \Bigl[ \sum_{\sigma} 
    \Bigl( - t_{ij} c_{j\sigma}^\dagger c_{i\sigma} + {\rm H.c.}
    \Bigr) 
    + J_{\alpha} {\bf S}_{i} \cdot {\bf S}_{j} 
   \Bigr] , 
\end{eqnarray}
where the site sum is taken over nearest neighbor pairs, and 
the label, $\alpha$=$x (y)$, denotes the leg (rung) direction.  
The exchange constant is $J_x = J$ along legs and $J_y $=$ J'$ 
along rungs.  The hopping integral has a finite phase 
due to the magnetic field, 
$t_{ij} $=$ t \exp [ -i e \int_{{\bf r}_i}^{{\bf r}_j} 
{\bf A} ({\bf r}) \cdot d{\bf r}]$, 
with the Landau gauge, ${\bf A} $=$ B ( -y, 0, 0 )$.  
The periodic (open) boundary conditions 
are used along the leg (rung) direction.  

The conductivity at temperature $T$ is defined by the Kubo formula as 
\begin{eqnarray}
  &&\sigma_{\alpha \beta} (\omega) = 
 { i e^2 \over V} {1 \over \omega + i \eta} 
  \Bigl( 
     - \bigl< K_\alpha \bigr> \delta_{\alpha \beta} \nonumber\\
   && \ \ 
     - { 1 \over Z} \sum_{n,m} 
     { e^{-E_m /T} - e^{-E_n / T} 
       \over 
       E_n - E_m - \omega - i \eta }  
     \bigl< n | J_\alpha | m \bigr> 
     \bigl< m | J_\beta | n \bigr> 
  \Bigr) , 
  \label{Kubo}
\end{eqnarray}
where $V$ is the number of sites, 
$Z$=${\rm Tr} \, e^{-{\cal H}/T}$, $| n \rangle$'s are 
eigenstates of ${\cal H}$ with energies $E_n$, 
$\langle K_\alpha \rangle$ is the thermal average of the 
kinetic energy along the $\alpha$-direction, and 
$J_\alpha$ is the paramagnetic current operator.  
$\eta$ is an adiabatic constant.  
The Hall coefficient is defined as 
\begin{equation}
  R_H (\omega ) = { 1 \over B } 
  { \sigma_{xy} ( \omega ) \over 
    \sigma_{xx} ( \omega ) \sigma_{yy} ( \omega ) 
     + \sigma_{xy}^2 ( \omega ) 
  } .  
  \label{hall}
\end{equation} 
Shastry et al.\ suggested its high-frequency limit 
as a measure of $T$-dependent effective 
carrier density \cite{shastry}
\begin{equation}
  R_H^* \equiv 
  \lim_{\omega \rightarrow \infty} R_H (\omega ) 
  = { V \over i B } 
  { \bigl\langle [ J_x , J_y ] \bigr\rangle 
    \over 
    \bigl\langle -K_x \bigr\rangle 
    \bigl\langle -K_y \bigr\rangle }.  
  \label{RHlim}
\end{equation}
We have numerically obtained all the eigenstates 
using the Householder method, and calculated the conductivity 
and the Hall coefficient using Eqs.\ (\ref{Kubo})-(\ref{RHlim}).  
The canonical ensemble is used to fix the electron number with 
all the subspaces of $S_{\rm tot}^z$ included.  
Spin gap, $\Delta_s$, is also calculated as the difference of the 
ground state energy between the subspaces of $S_{\rm tot}^z =0$ and 1.  

Let us start with the $\omega$-dependence of conductivity.  
Figure \ref{fig:cond} shows $\sigma_{xx} (\omega )$ 
at various $T$'s for three typical cases: 
(a) uniformly coupled ladder (UL), $J/t$=$J'/t$=0.3, 
(b) rung-dominant ladder (RL) which has larger rung couplings 
than leg couplings, $J'/t$=$20J/t$=6.0, 
and (c) dimerized ladder (DL) where each chain is dimerized, 
$J/t$=$J'/t$=0.3, and $J''/t$=1.0.
To characterize the dimerization, we use the notation 
$J$=$J_{2j}$$<$$J''$=$J_{2j+1}$.  
At high $T$'s above the temperature scale of $\Delta_s$, 
$\sigma_{xx}(\omega )$ has a broad incoherent part. 
At these temperatures, the conductivity is well fitted by 
the scaling function proposed by Jakli\v{c} and Prelov\v{s}ek for 
the 2D $t$-$J$ model \cite{jaklic2}
\begin{equation}
  \label{scale}
  \sigma_{xx} (\omega ) = {1 - e^{-\omega / T} \over \omega} C( \omega ).  
\end{equation}
Here $C(\omega )$ is a smooth universal function with a broad width, 
$\gamma_0$ of the order of several times of $t$.  In our case, this may be 
well approximated by $C (\omega ) \sim C_0(T)e^{-(\omega/\gamma_0)^2}
[1-e^{-(\gamma_0/T + T/\omega)}]$ for small $\eta$ with a rather 
weak $T$ dependence of $C_0(T)$. 
The details will be discussed elsewhere.  
The broad structure of $C(\omega)$ is a 
consequence of rapid and incoherent charge dynamics, since 
as shown from Eq. (\ref{Kubo}), $C(\omega )$ is the current-current 
correlation function, 
$C( \omega ) = \int dt \langle J (t ) J (0) \rangle e^{i \omega t}$.  
We note that this incoherent charge dynamics is consistent with
the anomalous suppression of coherence near the Mott insulator 
derived in the scaling theory \cite{imada}.
It should also be noted that 
this scaling function 
with weak $T, \omega $ dependence of $C(\omega)$
corresponds to the relaxation rate proposed in  
the marginal Fermi liquid theory \cite{mfl}
$\tau^{-1} \sim \max ( \omega , T )$ at low $\omega$ and $T$.  

Below a characteristic temperature, $T_{\rm PG}$, 
the conductivity $\sigma_{xx} (\omega )$ turns to show 
a pseudogap at $\omega$$\leq$$\omega_{\rm PG}$ as clearly shown 
in Fig.\ \ref{fig:cond} (b) and (c).  
This is accompanied by the appearance of a prominent coherent 
peak in the low-$\omega$ region, particularly for Fig.\ \ref{fig:cond} (b),  
and its weight grows with decreasing $T$.  By changing 
$J$ and $J'$, we have found that 
$T_{\rm PG}$$\simeq$$\Delta_s$$\simeq $$\omega_{\rm PG}$. 
Note that $\Delta_s =2.402$ for (b) and $\Delta_s =0.812$ for (c).
The magnetic susceptibility, $\chi (T)$, 
is in fact substantially reduced at $T < T_{\rm PG}$.  
In the case of Fig.\ \ref{fig:cond} (a),  
the pseudogap behavior is 
not prominent down to the lowest temperature used in the calculation
because of the small spin gap $\Delta_s /t \approx 0.002$.  

Comparing Fig.\ \ref{fig:cond} (b)(namely RL) and (c)(DL) at $T<T_{\rm PG}$, 
it is noticed that the RL has a larger 
coherent weight.  The observed $1/\omega^2$-behavior of 
$\sigma_{xx} (\omega )$ at $\omega /t$$\alt$1 in (b) is 
actually due to a finite $\eta$.  
The $\eta$-dependence at small $\eta$ indicates that  
this coherent peak is indeed a $\delta$-function, 
$\pi D^* \delta(\omega )$, within our 
numerical results, although $T >0$.  
In contrast, $\sigma_{xx} (\omega )$ for the DL in Fig.\ \ref{fig:cond} (c) 
shows a larger incoherent weight, 
even below $T_{\rm PG}$.  

The pseudogap behavior below $T_{\rm PG}$ 
indicates the formation of singlet  
bound pairs of electrons, which behave as singlet hard-core bosons.  
This is consistent with the exact diagonalization results 
at $T=0$ \cite{dagotto,tsune,hayward}.  The part of 
$ \omega$$ <$$ \Delta_s$ and $ T $$<$$ T_{\rm PG}$ is therefore described 
by an effective model of hard-core bosons with charge $+2e$, 
belonging to the universality class of Luther-Emery liquids \cite{LEL}.

The bound pairs are formed on a rung in the RL's
while on a dimerized strong bond in the DL's.
Therefore it leads to different effective models, i.e., hard-core 
bosons on a single {\em chain} and on a {\em ladder}, respectively, 
with a small nearest-neighbor hopping $t^*$ and interactions $V^*$ as 
calculated in Ref.\ \cite{tsune}.     
The boson ladder and chain show an incoherent behavior, 
above a characteristic temperature, $T_{\rm coh}$,
which is similar to the incoherent behavior of the 
Fermion ladder at $T$$\agt$$T_{\rm PG}$ discussed for Fig.\ \ref{fig:cond}
Whereas a dissipationless Drude peak, 
$\sigma_{xx} (\omega )$$ \sim$$ \pi D^* (T) \delta (\omega )$,
appears 
below $T_{\rm coh}$ with 
$D^* (T)$$ \approx $${\rm const}$.  
$T_{\rm coh}$ also corresponds 
to the crossover in the $T$-dependence of the kinetic energy:  
roughly speaking, $\langle K_x \rangle \propto 1/T$ above $T_{\rm coh}$ 
and $\langle K_x \rangle $$\approx$$ {\rm const.}$ below it.  
Based on a finite-size scaling of $\langle K_x \rangle (T)$, 
we found  $T_{\rm coh} $$\propto $$\delta^2 t^* $, 
consistent with the prediction of the scaling theory 
for the metal-insulator transition in 1D \cite{imada}.  

The $\sigma (\omega , T)$ of the boson chain was calculated by 
Zotos and Prelov\v{s}ek \cite{zotos}.  
They found that there remains a real $\delta$-function 
of the Drude peak even at temperatures of order $t^*$, 
and attributed it to the 
integrability of the model.  Our results are consistent with 
their observation.  

We have calculated the conductivity by the same numerical method 
for hard-core boson chains and ladders.  
The results show that 
the Drude weight for the boson ladders is smaller 
than for the boson chains, suggesting a more diffusive character 
of the conductivity, whereas it turns out that 
$T_{\rm coh}$ is higher for the ladders.  
The scaling theory for 
a single component system\cite{imada} predicted 
$D$$\propto$$\delta  $ and $T_{\rm coh}$$\propto$$\delta^2$ in 1D, and 
$D$$\propto$$\delta$ and $T_{\rm coh}$$\propto$$\delta$ in 2D.  
The difference between the results in 
Fig.\ \ref{fig:cond} (b) and (c) implies a crossover from 
1D to 2D.

Now let us discuss the Hall coefficient at $\omega \rightarrow \infty$, 
$R_H^*$.  Figure \ref{fig:RHlim} shows the $T$-dependence 
of $R_H^*$ for various ladders including those used for Fig.\ \ref{fig:cond}.    
In the limit of $T \rightarrow \infty$, $R_H^*$ is finite and positive 
in all cases.   For the 
$t$-$J$ ladder with hole doping $\delta$, it is indeed given by 
$  \lim_{T \rightarrow \infty} R_H^* = 
    -{\delta \over 1-\delta} + {1-\delta \over 4 \delta} 
$, 
identical to the result for the square lattice\cite{shastry}, where 
it is positive if $0 < \delta < {1 \over 3}$. 
With lowering $T$, $R_H^*$ decreases and 
becomes negative at a certain temperature, $T_{\rm AF}$, 
as in the case of the 2D Hubbard model \cite{assaad}.
It suggests the same origin of this behavior with the 2D case 
at $T$$<$$T_{\rm AF}$, i.e., the gradual formation of electron-like 
large ``Fermi surface'' in the presence of two 
spin species.  

In the UL's ($J/t$=$J'/t$=0.3 and 1.5), 
$R_H^*$ remains negative at $T$$<$$T_{\rm AF}$. With decreasing 
$T$, it shows a minimum and finally converges to a large negative 
value at $T$=0.  
In the RL's ($J/t$=0.3, $J'/t$=6.0 and 20.0), 
$R_H^*$ decreases with lowering $T$ as in the previous case, 
but there appears a spike-like structure at around $T_{\rm PG}$.  
This structure is more prominent for larger $J'$'s.  
In the DL's, $R_H^*$$<$0 in a very narrow $T$-region and 
$R_H^*$ increases to positive and converges to a large positive 
value at $T$=0.  

The low-temperature behaviors, particularly 
at $T$$ <$$ T_{\rm PG}$, depend on the coupling constants, 
meaning distinct characters of charge carriers 
for the different cases.  
Since $\langle -K_\alpha \rangle $$>$0 at any $T$'s, 
the sign of $R_H^*$ is determined by the sign of 
$\langle [J_x , J_y ] \rangle$, i.e., that of $\sigma_{xy}$.  
Considering the formation of bound hole pairs below $T_{\rm PG}$,
it is natural to assume that $\sigma_{xy}$ in 
Eq.\ (\ref{hall}) is obtained as the sum of 
contributions from bound pairs and unpaired electrons.  
The $T$-dependence of 
$R_H^*$ is therefore qualitatively explained from $T$-dependence 
of the density and mobility of these two types of carriers.  

As for the unpaired electrons, their density decreases 
exponentially at $T$$ <$$ T_{\rm PG}$.  Equation (\ref{RHlim}) 
shows their Hall mobility is determined by the processes  
of moving an unpaired electron first along one direction 
and then perpendicular to it.  
It is important to notice that the effective hopping matrix elements 
of unpaired electrons acquire an extra Fermion negative sign when 
they undergo a recombination scattering with a bound pair.  
Therefore, the Hall mobility 
of unpaired electrons is positive at $T$$ >$$ T_{\rm AF}$ and 
$T$$ \ll $$T_{\rm PG}$ while negative at 
$ T_{\rm PG}$$<$$ T $$<$$ T_{\rm AF}$, 
changing its sign twice.  

When the bound pairs are concerned, we recall that 
$\sigma_{xy}$ at $T \ll T_{\rm PG}$ is mainly determined by this part, 
since only few unpaired electrons are thermally excited.  
The bound pairs are formed on the rungs in the RL's, 
whereas within the legs for the DL's.  
The opposite sign of the Hall mobility at $T < T_{PG}$ 
between these two cases may be attributed to 
these different local configurations of bound pairs, which will 
be discussed elsewhere.

The Hall coefficient for small $\omega$ is shown in Fig.\ \ref{fig:RHzero} 
for the same ladders in Fig.\ \ref{fig:RHlim}.  
The $T$-dependence is qualitatively similar 
to $R_H^*$ for the UL's and the DL's.  
On the other hand, the RL's 
have a different $T$-dependence at small $\omega$'s.  
This may be due to finite size corrections, since 
$R_H$ at small $\omega$'s is very sensitive to the value 
of the adiabatic constant $\eta$.  Absolute value of $R_H$ 
in Fig.\ \ref{fig:RHzero} (a)-(d) also changes considerably with 
varying $\eta$.  

Figure \ref{fig:RHdens} shows the hole doping dependence of 
$R_H^*$ at $T$=0 calculated by Lanczos diagonalization 
for the UL's.  
When the electron density is small, $R_H^* < 0$, meaning 
electron-like.  
When $\delta \ge { 1 \over 2}$, 
the value of $R_H^*$ is close the classical value, i.e., 
$ R_H^{\rm class}$=$ -{ 1 \over 1-\delta }$$<$
$ R_H^* (\mbox{$T$=0})$$<$$ R_H^* ( \mbox{$T$=$\infty$} )$$<$0, 
showing little $J$-dependence. The Hall coefficient in this region 
is mainly determined by the electron density. 
On the other hand, at small $\delta$'s, $R_H^*$ 
shows a substantial $J$-dependence, and when $J/t$$ >$1 it is 
rather close to $1/\delta$ except for its negative sign.  
This means the importance of electron correlation effects 
and is consistent with the previous conclusion that the 
bound pairs are effective charge carriers.  A comment is 
necessary for the positive sign of $R_H^*$ at $\delta$=${1 \over 8}$ 
and $J/t$$<$0.2.  This may be due to strong ferromagnetic fluctuations.  
As shown in Ref.\ \cite{tsune}, the Nagaoka ferromagnetic 
ground state appears near half filling at small $J$'s.  
The ``Fermi volume'' of the electrons with the majority spin 
is doubled, leading to a hole-like $R_H^*$.  

We finally discuss on how the Hall 
coefficient is affected when ladders are coupled.  
Figure \ref{fig:RHcouple} shows $R_H^*(\mbox{$T$=0})$ 
for various {\em coupled ladders} calculated by 
Lanczos diagonalization using Eq.\ (\ref{RHlim}).  
Two two-leg ladders are coupled by interladder hopping 
$t_L$ and exchange $J_L$, to form a 4$\times$4-site cluster.  
The finite coupling of the ladders shows shifts of 
$R_H^*$ toward the positive direction in general.  
A typical case is weakly coupled DL's and its equivalent configurations 
(for example, 
$J''$$>$$J$=$J'$=10$J_L$=0.3$t$ or 
0.03$J'$$>$$J$=$J''$=10$J_L$=0.3$t$ or 
0.6$J'$$>$$J$=$J''$=$J_L$=0.3$t$).
A few exceptions with the negative $R_H^*$ are the cases of 
(i) weakly coupled RL's, 
(ii) the configuration of weakly coupled plaquettes
(e.g., the coupled DL at $J_L$$ < $$J$$ <$$ J''$$\leq $$J'$).
The crossover from a small to a large positive $R_H$ at $T_{\em PG}$ in 2D 
configurations is reminescient of the crossover in the 
cuprates at $T=T_{\em cr}$, suggesting that it shares the same mechanism
due to preformed pairs below $T_{\rm PG}$ \cite{imada}.

In summary, the frequency-dependent conductivity and Hall coefficient
of various types of doped ladders were calculated by numerical 
diagonalization.  There is a crossover from 
high-$T$ incoherent to low-$T$ coherent dynamics 
with the appearance of the Drude peak.  
Its crossover temperature $T_{\rm coh}$ is enhanced by the 
spin gap and resultant 
pseudogap formations. 
$T_{\rm coh}$ increases with increasing dimensionality 
when the spin gap is formed, although the coherent weight itself decreases.
The Hall coefficient shows a strong crossover to large amplitudes at low 
temperatures below the spin-gap temperature.    

The authors thank Fakher Assaad for fruitful discussions.  
This work was supported by a Grant-in Aid for Scientific Research 
from the Ministry of Education, Science, and Culture 
of Japan.  The numerical calculations were mainly performed on 
VPP500 at the Institute of Solid State Physics, Tokyo University.

%%%%%%%%
%\psfull
\begin{figure}[hbt]
%\begin{center}
%  \leavevmode
%  \psfig{figure=fig1.eps,height=6cm}
%\end{center}
\caption{Frequency dependence of the conductivity of the $t$-$J$ 
ladder for various choices of parameters.  
(a) $J$=$J'$=0.3$t$, 2 holes in 2$\times$6 sites, 
(b) $J$=0.3$t$, $J'$=6.0$t$, 2 holes in 2$\times$6 sites, and 
(c) $J$=$J'$=0.3$t$, $J''$=$t$, 2 holes in 2$\times$4 sites.  
The adiabatic constant is (a)-(b):$\eta$=0.1$t$, (c): 0.01$t$.
(d) $I(\omega) = \int_0^\omega C(\omega ') d\omega '$ 
calculated for (a).  
}
\label{fig:cond} 
\end{figure}
%%%%%%%%

%%%%%%%%
\begin{figure}[hbt]
%\begin{center}
%  \leavevmode
%  \psfig{figure=fig2.eps,width=7.5cm}
%\end{center}
\caption{Temperature dependence of the high-frequency 
Hall coefficient of the $t$-$J$ ladder. 
The system size is (a)-(b): 2 holes in 2$\times$6 sites, 
and (c) 2 holes in 2$\times$4 sites.  
}
\label{fig:RHlim}
\end{figure}
%%%%%%%%

%%%%%%%%
\begin{figure}
%\begin{center}
%  \leavevmode
%  \psfig{figure=fig3.eps,width=7.5cm}
%\end{center}
\caption{Temperature dependence of the low-frequency 
Hall coefficient of the $t$-$J$ ladder. 
The system size is (a)-(b): 2 holes in 2$\times$6 sites 
($\delta = {1 \over 6}$), 
and (e) 2 holes in 2$\times$4 sites ($\delta = {1 \over 4}$). 
The adiabatic constant is $\eta =0.1t$.  
}
\label{fig:RHzero}
\end{figure}
%%%%%%%%

%%%%%%%%
\begin{figure}
%\begin{center}
%  \leavevmode
%  \psfig{figure=fig4.eps,width=6.5cm}
%\end{center}
\caption{
Electron-density dependence of $R_H^*$ for 
the $t$-$J$ ladder with uniform couplings, $J$=$J'$.  
2$\times$8 sites.  
}
\label{fig:RHdens}
\end{figure}
%%%%%%%%

%%%%%%%%
\begin{figure}
%\begin{center}
%  \leavevmode
%  \psfig{figure=fig5.eps,width=7cm}
%\end{center}
\caption{Hall coefficient of coupled $t$-$J$ ladders 
in the $\omega$=$\infty$ limit, $R_H^*$.  Two ladders with 4$\times$2 sites 
are coupled by interladder hopping $t_L$ and exchange $J_L$.  
}
\label{fig:RHcouple}
\end{figure}
%%%%%%%%
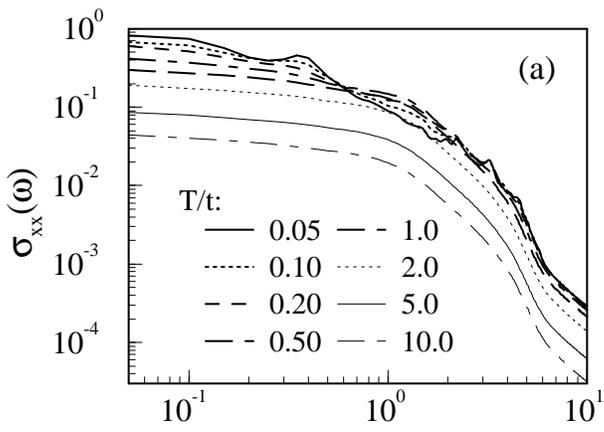
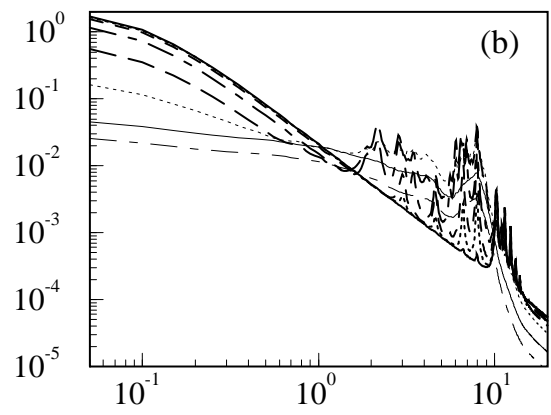
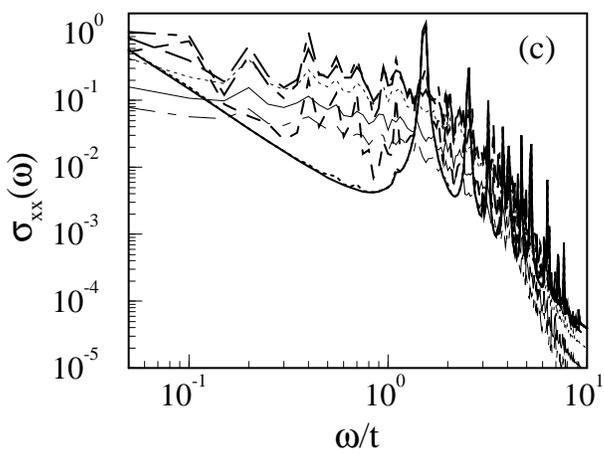
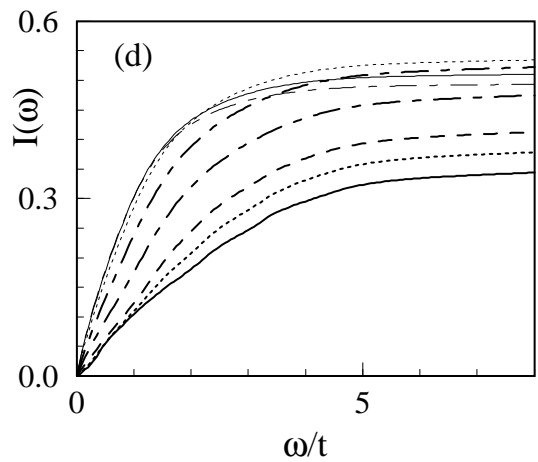
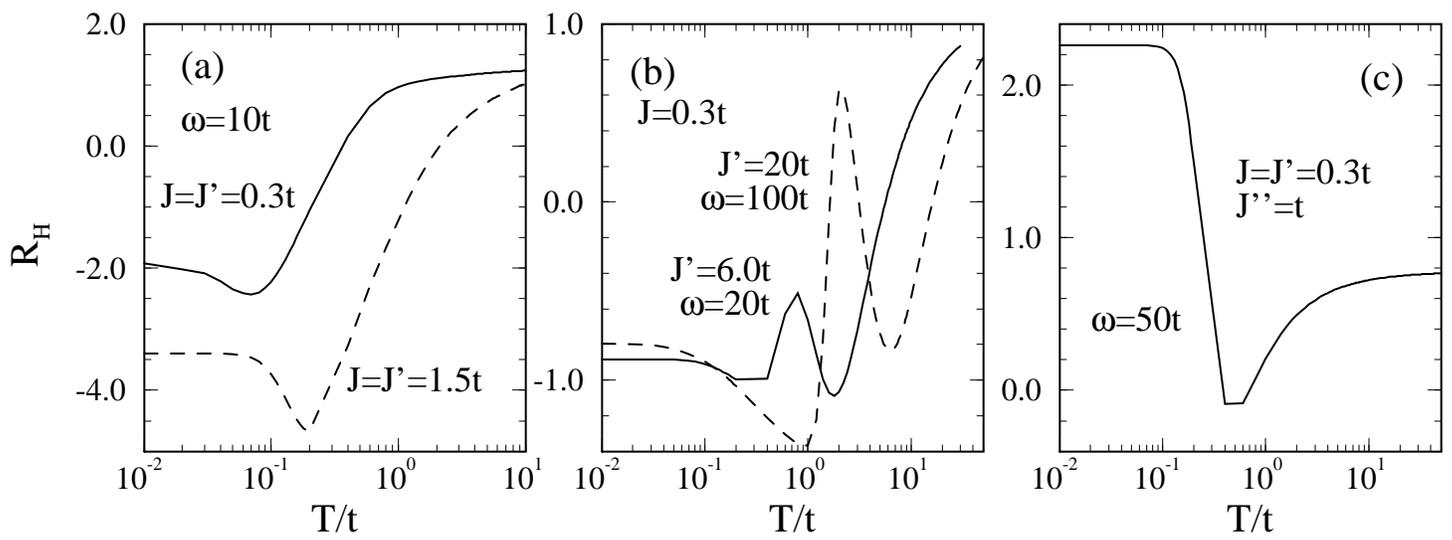
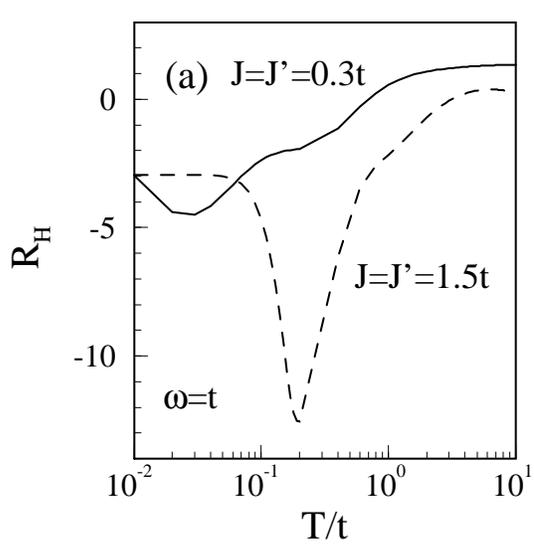
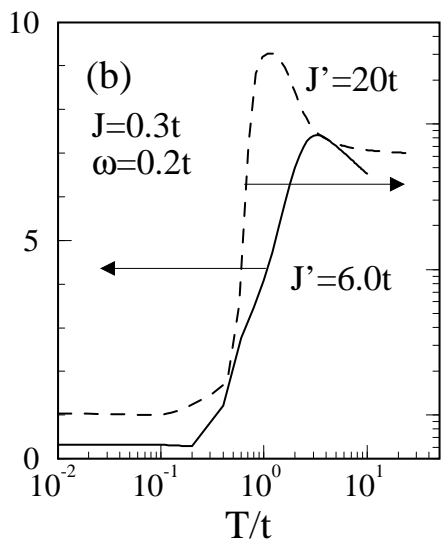
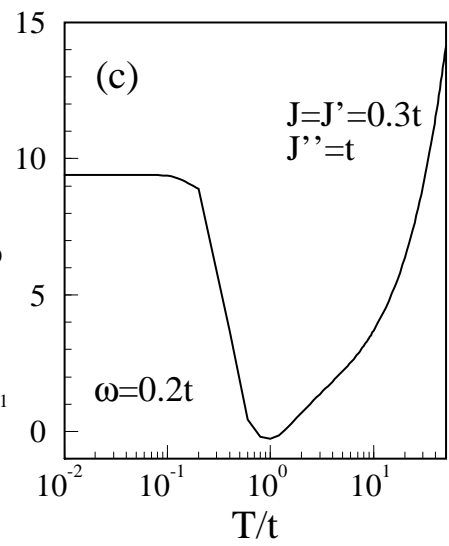
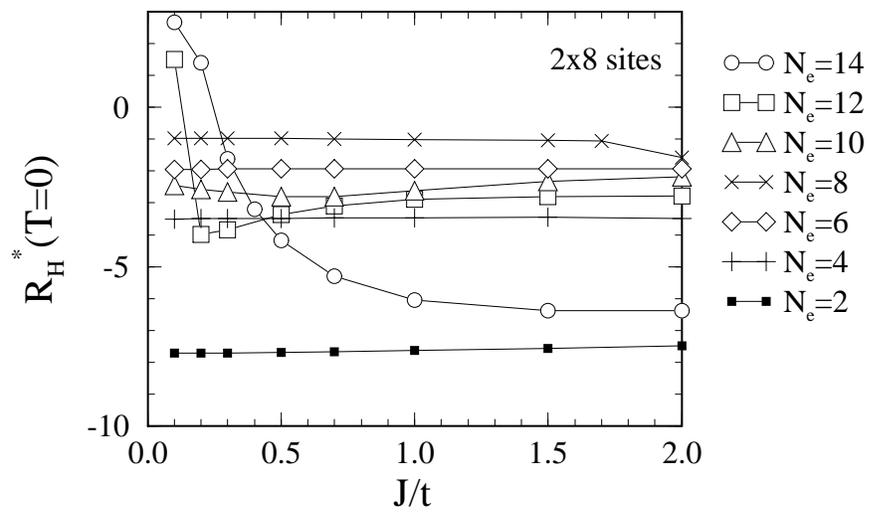
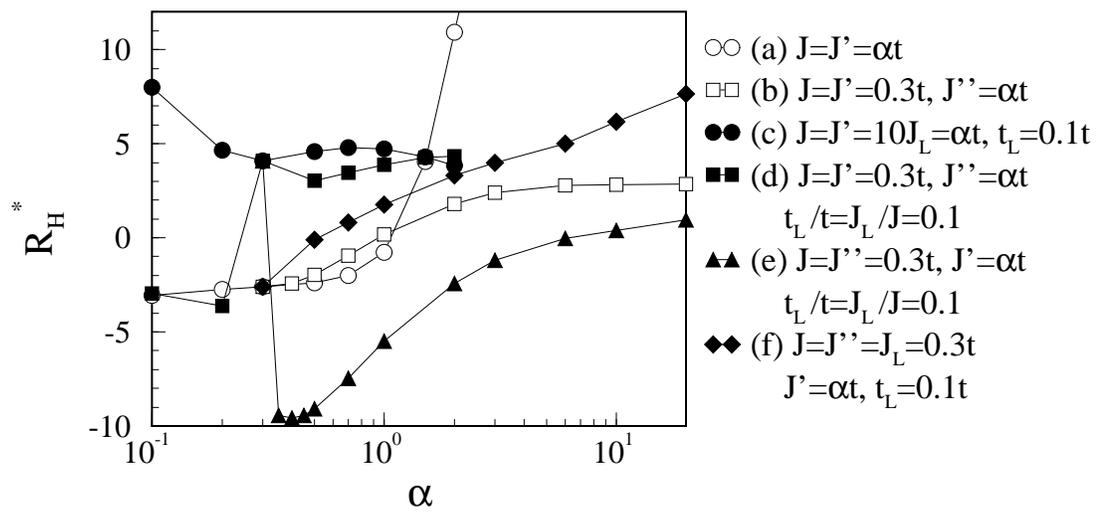
\end{document}